\documentclass[onecolumn,showpacs,amsmath,amssymb,aps,showkeys,floatfix,prd,a4paper]{revtex4}

\usepackage[dvips]{graphicx}
\usepackage{dcolumn}
\usepackage{bm}
\usepackage{epsfig}
\usepackage{amsfonts}
\usepackage{amssymb,amscd}

\def\lsim{\raise0.3ex\hbox{$<$\kern-0.75em\raise-1.1ex\hbox{$\sim$}}}
\def\gsim{\raise0.3ex\hbox{$>$\kern-0.75em\raise-1.1ex\hbox{$\sim$}}}

\def\pom{{I\!\!P}}

\def\beq{\begin{equation}}
\def\eeq{\end{equation}}
\def\bea{\begin{eqnarray}}
\def\eea{\end{eqnarray}}
\def\bq{\begin{quote}}
\def\eq{\end{quote}}

\newcommand{\rqq}{\mbox{\boldmath $q$}}

\def\gappeq{\mathrel{\rlap {\raise.5ex\hbox{$>$}}
{\lower.5ex\hbox{$\sim$}}}}

\def\lappeq{\mathrel{\rlap{\raise.5ex\hbox{$<$}}
{\lower.5ex\hbox{$\sim$}}}}

\def\Toprel#1\over#2{\mathrel{\mathop{#2}\limits^{#1}}}

\newcommand{\rk}{\mbox{\boldmath $k$}}

\def\pom{{I\!\!P}}

\def\imag{{\mathcal{I}\mathrm{m}}\,}

\begin{document}


\title{Double vector meson production from the BFKL equation}

\author{V.~P. Gon\c{c}alves}
\email{barros@ufpel.edu.br}
\affiliation{High and Medium Energy Group (GAME) \\
Instituto de F\'{\i}sica e Matem\'atica, Universidade Federal de Pelotas\\
Caixa Postal 354, CEP 96010-900, Pelotas, RS, Brazil}
\author{W.~K. Sauter}
\email{sauter@if.ufrgs.br}
\affiliation{High Energy Physics Phenomenology Group, GFPAE, IF-UFRGS \\
Caixa Postal 15051, CEP 91501-970, Porto Alegre, RS, Brazil}

\date{\today}

\begin{abstract}
The double vector  meson production in two photon collisions is addressed assuming  that the color singlet $t$-channel exchange carries large momentum transfer.  We consider the non-forward solution of the BFKL equation  at high energy and large momentum transfer and  estimate the total and differential cross section for the process $\gamma \gamma \rightarrow V_1 V_2$, where $V_1$ and $V_2$ can be any two vector mesons  ($V_i = \rho, \omega, \phi, J/\Psi, \Upsilon$).  A comparison between our predictions and previous theoretical results obtained at Born level or assuming  the Pomeron-exchange factorization relations is presented. Our results demonstrate that the BFKL dynamics implies an enhancement of the cross sections. Predictions for the future linear colliders (TESLA, CLIC and ILC) are given. 

\end{abstract}


\maketitle

\section{Introduction}
\label{intro}

Understanding the behavior of high energy hadron reactions from a fundamental perspective within Quantum Chromodynamics (QCD) is an important goal of particle physics. Attempts to test experimentally this sector of QCD have started some years ago  with the first experimental results from lepton - hadron (DESY-HERA) and hadron - hadron (FERMILAB-TEVATRON) collisions. On the other hand, the papers which form the core of our knowledge of the Regge limit (high energy limit) of QCD were established in the late 1970s by Lipatov and collaborators \cite{BFKL}. The physical effect that they describe is often referred to as the QCD Pomeron, with the evolution described by the BFKL equation. The simplest process where  this equation applies  is the high energy scattering between two heavy quark-antiquark states, {\it i.e.} the onium-onium  scattering. For a sufficiently heavy onium state, high energy scattering is  a perturbative process since the onium radius gives the essential scale at which the running coupling $\alpha_s$ is evaluated. This process was proposed as a {\it gedanken} experiment to investigate the high energy regime of QCD in Refs. \cite{mueller,muepat,muechen} (see also Refs. \cite{nik}).  At leading order (LO),  the cross section grows rapidly with the energy, 
\[\sigma \propto \alpha_s^2  \, e^{(\alpha_{\pom} - 1)Y},\;\mathrm{with}\quad \alpha_{\pom} - 1 = \frac{4\alpha_s\,N_c}{\pi}\ln 2 \quad\mathrm{and}\quad Y =\ln\,\frac{s}{Q^2},\]
 because the number of dipoles in the light cone wave function grows rapidly with the energy. Some examples of processes which are directly related with the onium - onium scattering are forward jets in deeply inelastic events at low values of the Bjorken variable $x$ in lepton-hadron scattering, jet production at large rapidity separations in hadron-hadron collisions and off-shell photon scattering at high energy in $e^+\,e^-$ colliders, where the photons are produced from the leptons beams by bremsstrahlung (For a review of BFKL searches, see e.g. Ref. \cite{reviewhighenergy}). This last process presents some theoretical advantages as a probe of QCD Pomeron dynamics compared to the other ones because it does not involve a non-perturbative target \cite{gamagama,gambrod,gamboone}. In particular, the  photon colliders offer unique possibility to probe QCD in its high energy limit \cite{kw}. The simplicity of the initial state and the possibility of study of many different combinations of final states making this process very useful for studying the QCD dynamics in  the limit of high center-of-mass energy $\sqrt{s}$  and fixed momentum transfer $t$. For instance, if we consider quark production in $\gamma \gamma$ collisions we have that in the general case there can be three  large momentum scales in this process - the photon virtuality $Q^2$, the  momentum transfer $t$ and the  quark mass $M$. It implies that for real photons interactions and $t = 0$ only heavy quark production can be treated perturbatively. On the other hand, perturbation theory can be used even for light quark production if $|t| \gg \Lambda_{QCD}^2$. In  the last years many authors have studied in detail the heavy quark production in $\gamma \gamma$ collisions considering different theoretical approaches \cite{mottim,hansson} (See also Ref. \cite{per2}). 

Other possibility for the study of the QCD Pomeron is the vector meson pairs production in $\gamma \gamma$ collisions \cite{ginzburg}. At very high energies $s \gg  -t$, diffractive processes such as $\gamma \gamma \rightarrow $ neutral vector (or pseudo-scalar) meson pairs with virtual or real photons can test the QCD Pomeron (Odderon) in a detailed way utilizing the simplest possible initial state. As in the case of the large angle exclusive $\gamma \gamma$ processes, the scattering amplitude is computed by convoluting the hard scattering pQCD amplitude for $\gamma \gamma \rightarrow q\overline{q} q\overline{q}$ with the vector meson wave functions. For heavy vector mesons, this cross section can be calculated using the perturbative QCD methods. First calculations considering the Born two-gluon approximation have been done in Refs. \cite{ginzburg}. Some years ago, the double $J/\Psi$ production in $\gamma \gamma$ collisions has been proposed as a probe of the QCD Pomeron \cite{motyka}, where the cross sections are calculated in the BFKL framework and also considering a next leading order (NLO) calculation by a kinematic veto on the gluon momenta in the ladder. Non-perturbative contributions for this process are studied in \cite{DucatiSauter} by employing modified gluon propagators. Theoretical estimates of the cross sections presented in \cite{motyka} have demonstrated that measurement of this reaction at a Photon Collider should be feasible (See also Ref. \cite{per1}). On the other hand, in order to calculate light vector meson production it is necessary to consider virtual photon collisions or a large momentum transfer. In this paper we study the double vector meson production in $\gamma \gamma$ collisions, where the the color singlet $t$-channel exchange carries large momentum transfer.   Such processes are characterized by a large rapidity gap between the vector mesons, with it identified experimentally by their decay into charged particles, {\it e.g.} $\rho \rightarrow \pi^+ \pi^-$. The presence of a large momentum transfer allow us to calculate the light and heavy vector mesons production cross sections for real photon interactions. We consider the non-forward  solution of the BFKL equation  at high energy and large momentum transfer and  estimate the total and differential cross section for the process $\gamma \gamma \rightarrow V_1 V_2$, where $V_1$ and $V_2$ can be any two vector mesons  ($V_i = \rho, \omega, \phi, J/\Psi, \Upsilon$). Our study is motivated by the fact that the experimental data for the vector meson photo-production at high $t$ in electron-proton collisions at HERA can be described quite well using the impact factor representation and the  non-forward BFKL solution \cite{FP,jhep}. In order to simplify our calculations we will consider the non-relativistic approximation in the calculations of the impact factors. This approximation should be not applicable for light mesons. However, as verified in Ref. \cite{FP}, the predictions obtained using the non-forward BFKL solution  at asymptotic energies (leading conformal spin solution) and this approximation agree quite well  with heavy and light meson HERA data. This phenomenological success occurs due to the selection of the dominant amplitude of the general expression as demonstrated  in Ref. \cite{jhep}. We will assume that this dominance will be also present in the processes considered in this paper, and use this approximation in that follows. We postponed for a future publication the study of double vector production using the full amplitude, similarly to Ref. \cite{jhep},  which will allow us consider the helicity flip of the quarks. 

This paper is organized as follows. In next section we present a brief review of the formalism necessary in order to calculate the differential and total cross section of the $\gamma \gamma \rightarrow V_1 V_2$ process. In Section \ref{results} we present our predictions and compare our results with previous estimatives. Finally, in Section \ref{conc} we present a summary of our main conclusions.

\section{Formalism}
\label{form}

\begin{figure}[t]
\centerline{\psfig{file=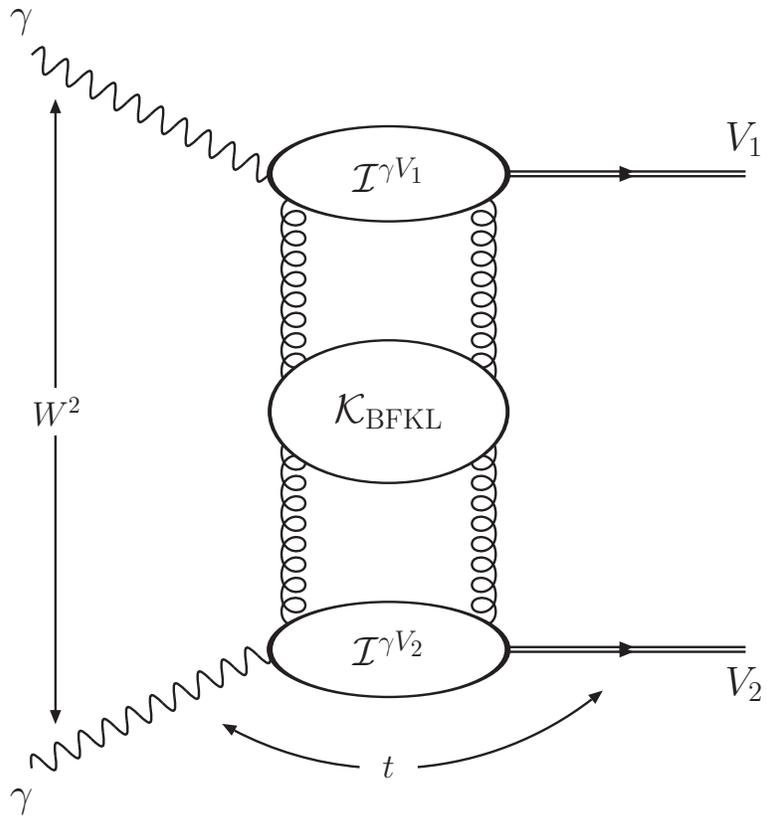,width=100mm}}
 \caption{Impact factor representation for the $\gamma \gamma \rightarrow V_1 V_2$ process.}
\label{fig1}
\end{figure}

In order to calculate the $\gamma \gamma \rightarrow V_1 V_2$ cross section we  will consider the impact factor representation proposed by Cheng and Wu \cite{ChengWu} many years ago, which allow us factorize the process in three disjoint parts: the impact factors associated to the transitions ${\mathcal{I}}^{\gamma V_i} \,\,\,(i = 1,2)$ and the exchange mechanism between them represented by ${\cal K}_{BFKL}$ in Fig. \ref{fig1}. In this representation, the amplitude for the high energy process $A B\to C D$ can be expressed on the form 
\begin{equation}
{\cal A}^{A B\to C D} (\rqq) = 
\int {d^2 \rk \, d^2\rk^{\prime} }\,\,
 {\mathcal{I}}^{A\to C}(\rk, \rqq)\,\, \frac{{\cal{K}}_{BFKL}(\rk,\rk^{\prime},\rqq)}{\rk ^2 (\rqq - \rk)^2} \,\,{\mathcal{I}}^{B\to D} (\rk^{\prime}, \rqq),
\label{mpr}
\end{equation}
where ${\mathcal{I}}^{A\to C}$ and ${\mathcal{I}}^{B\to D}$ are the impact factors for the upper and lower parts of the diagram, respectively.\ That is, they are the impact factors for the processes $A\to C$ and $B\to D$ with two gluons carrying transverse momenta $\rk$ and $\rqq-\rk$ attached, the gluons being in an overall color singlet state. At lowest order the process is described by two gluon exchange, which implies ${\cal K}_{BFKL} \propto \delta^{(2)} (\rk - \rk^{\prime}) $ and an energy independent cross section. At higher order, the dominant contribution is given by the QCD pomeron singularity which is generated by the ladder diagrams with the (reggeized) gluon exchange along the ladder. The QCD pomeron is described by the BFKL equation \cite{BFKL}, which implies that the exchange of the gluon ladder with interacting gluons generates increasing cross sections with the energy. As our goal  is the analysis of double vector meson production at large - $t$, we will use in our calculations the non-forward solution of the BFKL equation in the leading logarithmic approximation, obtained by Lipatov in Ref. \cite{Lipatov}. 

The differential cross-section, characterized by the invariant collision energy squared $s$, is expressed in terms of the amplitude  ${\mathcal{A}}(s,t)$ as follows
\begin{equation}
\frac{d \sigma}{dt} = \frac{1}{16 \pi} |{\mathcal{A}}(s,t)|^2.
\label{dsdt}
\end{equation}
The amplitude is dominated by its imaginary part, which we shall parametrize, as in \cite{FP,FR}, by a dimensionless quantity  $\mathcal{F}$
\begin{equation}
\imag {\mathcal{A}}(s,t) =
\frac{16 \pi}
{9 t^2} {\mathcal{F}}(z,\tau)
\label{FA}
\end{equation}
where $z$ and $\tau$ are defined by
\begin{equation}
z = \frac{3\alpha_{s}}{2\pi} \ln \biggl( \frac{ s}{\Lambda^{2}} \biggr)
\label{zdef}
\end{equation}
\begin{equation}
\tau = \frac{|t|}{M_{V}^{2}+ Q_{\gamma}^{2}},
\label{taudef}
\end{equation}
where
$M_{V}$ is the mass of the vector meson, $Q_\gamma$ is the photon virtuality and $\Lambda^{2}$ is a characteristic  scale related to $M_V^2$  and $|t|$. 
In this paper we only consider $Q_\gamma=0$. In LLA, \( \Lambda  \) is arbitrary (but must depend on the scale in the problem, see discussion below) and \( \alpha _{s} \) is a constant. For completeness, we give the cross-section expressed in terms of ${\mathcal{F}}(z,\tau)$, where the real part of the amplitude is neglected,
\begin{equation}
\frac{d\sigma (\gamma \gamma \rightarrow V_1 V_2)}{dt} \; = \;
\frac{16\pi}{81 t^4}
|{\mathcal{F}}(z,\tau)|^{2}.
\label{dsdtgq}
\end{equation}
This representation is rather convenient for the calculations performed below.

The BFKL amplitude, in the leading logarithm approximation (LLA) and lowest conformal spin ($n=0$), is given by~\cite{Lipatov}
\begin{equation}
\label{BFKLa}
{\mathcal{F}}_{\mathrm{BFKL}}(z,\tau)=\frac{t^{2}}{(2 \pi)^{3}}\int d\nu \frac{\nu ^{2}}{(\nu ^{2}+1/4)^{2}}e^{\chi (\nu )z}I_{\nu }^{\gamma V_1}(Q_{\perp })I^{\gamma V_2}_{\nu }(Q_{\perp })^{\ast },
\end{equation}
 where $Q_{\perp}$ is the momentum transfered, $t=-Q_{\perp}^2$, (the subscript denotes two dimensional transverse vectors) and 
\begin{equation}
\chi (\nu )=4{\mathcal{R}}\mathrm{e}\biggl (\psi (1)-\psi \bigg (\frac{1}{2}+i\nu \bigg )\biggr )
\end{equation}
is proportional to the BKFL kernel eigenvalues~\cite{Jeff-book}, with $\psi(x)$ being the digamma function. 

 The quantities $I_{\nu }^{\gamma V_i}$ are given in terms of the impact factors ${\mathcal{I}}_{\gamma V_i}$ and the BFKL eigenfunctions as follows \cite{FR},
\begin{eqnarray}
I_{\nu}^{\gamma V_i}(Q_{\perp }) & = & \int \frac{d^{2}k_{\perp }}{(2\pi )^{2}}\, {\mathcal{I}}_{\gamma V_i}(k_{\perp },Q_{\perp })\int d^{2}\rho _{1}d^{2}\rho _{2} \label{inugv}\\
 & \times  & \biggl [\left( \frac{(\rho _{1}-\rho _{2})^{2}}{\rho _{1}^{2}\rho _{2}^{2}}\right) ^{1/2+i\nu }-\left( \frac{1}{\rho _{1}^{2}}\right) ^{1/2+i\nu }-\left( \frac{1}{\rho _{2}^{2}}\right) ^{1/2+i\nu }\biggr ]e^{ik_{\perp }\cdot \rho _{1}+i(Q_{\perp }-k_{\perp })\cdot \rho _{2}}.\nonumber 
\end{eqnarray}
In the case of coupling to a colorless state only the first term in the square bracket survives since \( {\mathcal{I}}_{A}(k_{\perp },Q_{\perp }=k_{\perp })={\mathcal{I}}_{A}(k_{\perp }=0,Q_{\perp })=0 \) in this case. 
The impact factor ${\mathcal{I}}_{\gamma V_i}$ describes in the high energy limit the couplings of the external particle pair to the color singlet gluonic ladder. It are obtained in the perturbative QCD framework and we approximate them by the leading terms in the perturbative expansion \cite{Ryskin}:
\begin{eqnarray}
{\mathcal{I}}_{\gamma V_i}\;=\;
\frac{{\mathcal C}_i \alpha_s}{2}\, \biggl(\frac{1}{\bar{q}^{2}}-
\frac{1}{q_{\|}^{2}+k_{\perp}^{2}} \biggr).
\label{impfmom}
\end{eqnarray}
In this formula, factorization of the scattering process and the meson formation is assumed, and the non-relativistic approximation of the meson wave function is used. In this approximation the quarks in the meson have collinear four-momenta and $M_{V}=2M_{q}$ where $M_{q}$ is the mass of the constituent quark. To leading order accuracy, the constant ${\mathcal C}_i$ may be related to the vector meson leptonic decay width
\begin{eqnarray}
\mathcal{C}^{2}_i\;=\;\frac{3\Gamma_{ee}^{V_i}M_{V_i}^{3}}{\alpha_{\mathrm{em}}}.
\end{eqnarray}
Moreover, we have that 
\begin{eqnarray}
\bar{q}^{2}\;=\;q_{\|}^{2}+Q_{\perp}^{2}/4, \\
q_{\|}^{2}\;=\;(Q_{\gamma}^{2}+M_{V_i}^{2})/4.
\label{C}
\end{eqnarray}
Although this approximation should be not applicable for light mesons we will use it here following  \cite{FP}, which has observed that  this approximation 
implies a reasonable description of the heavy and light meson HERA data. As emphasized in Ref. \cite{jhep}, this phenomenological success occurs due to the selection of the dominant amplitude of the general expression. We will assume that this dominance will be also present in the processes considered in this paper. We postponed for a future publication the study of double vector production using the full amplitude (also considering higher conformal spins contributions).

Using Eq. (\ref{impfmom}) into (\ref{inugv}), one obtains \cite{FR,BFLW}
\begin{eqnarray}
I_{\nu}^{\gamma V_i}(Q_{\perp }) & = & -{\mathcal{C}_i}\, \alpha_s \frac{16\pi}{Q_{\perp }^{3}}\frac{\Gamma (1/2-i\nu )}{\Gamma (1/2+i\nu )}\biggl (\frac{Q_{\perp }^{2}}{4}\biggr )^{i\nu }\int _{1/2-i\infty }^{1/2+i\infty }\frac{du}{2\pi i}\biggl (\frac{Q_{\perp }^2}{4 M_{V_i}^2}\biggr )^{1/2+u}\\
 &  & \times\frac{\Gamma ^{2}(1/2+u)\Gamma (1/2-u/2-i\nu/2)\Gamma (1/2-u/2+i\nu/2)}{\Gamma (1/2+u/2-i\nu /2)\Gamma (1/2+u/2+i\nu /2)}.\nonumber \label{IV} 
\end{eqnarray}
The differential cross section can be directly calculated substituting the above expression in Eq. (\ref{BFKLa}) and evaluating numerically the integrals. The total cross section will be given by
\begin{eqnarray}
\sigma (\gamma \gamma \rightarrow V_1 V_2) = \int_{|t|_{min}}^{\infty} d|t| \,\,\, \frac{d\sigma (\gamma \gamma \rightarrow V_1 V_2)}{d|t|} \,\,,
\label{sigtot}
\end{eqnarray}
where $|t|_{min}$ is the minimum momentum transfer, which we will assume different values for the distinct processes.

\section{Results and discussions}
\label{results}

\begin{figure}[t]
\centerline{\psfig{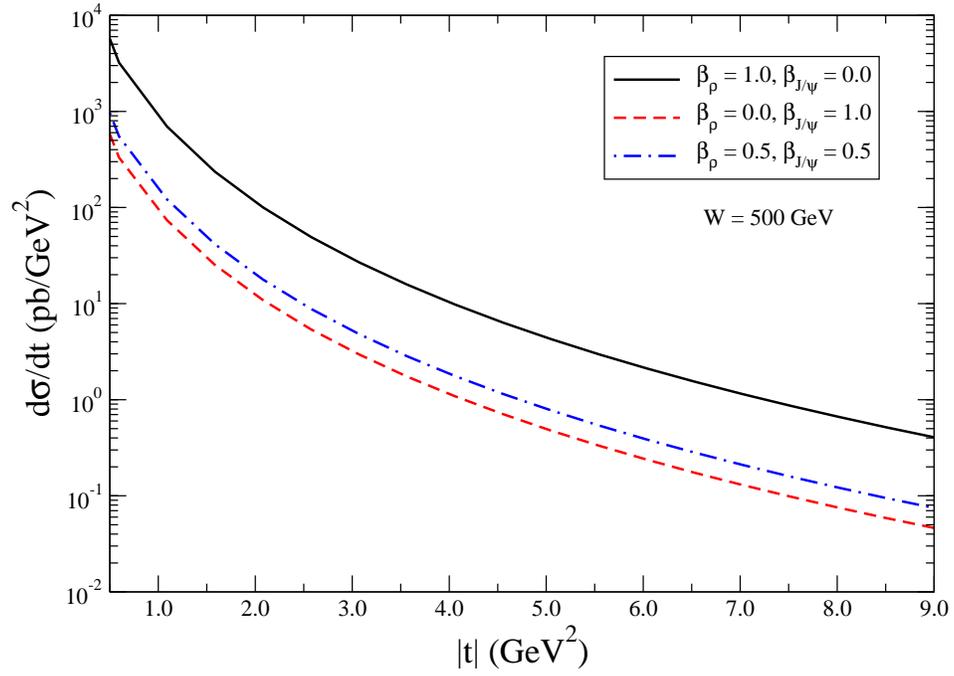}}
 \caption{Dependence of the differential cross section for $\rho J/\Psi$ production in the choice for the $\beta$ parameters. }
\label{fig2}
\end{figure}

Lets start the analysis of our results discussing the choice for the parameters $\alpha_s$ and $\Lambda$. We have that $\alpha_s$ appears in two different places in our calculations: in Eq. (\ref{IV}) and in the definition of the variable $z$ [Eq. (\ref{zdef})]. The first factor comes from the couplings of the two gluons to the impact factor and the second one is generated by the gluon coupling inside the gluon ladder (For details see Ref. \cite{jhep}) . In this paper  we treat these strong couplings as being identical and assume a fixed $\alpha_s$, which is appropriate to the leading logarithmic accuracy. Furthermore, as the cross section is proportional to $\alpha_s^4$, our results are strongly dependent on the choice for the $\alpha_s$ value. We assume  $\alpha_s = 0.2$, following Refs. \cite{FP,jhep}, where this value was determined from  a fit to the HERA data (For a detailed discussion see Ref. \cite{FP}).
Similarly,  at leading logarithmic approximation,  $\Lambda$  is arbitrary but must depend on the scale in the problem. In our case we have that in general $\Lambda$ will be a function of $M_V$ and/or $t$. Following Ref. \cite{FP} we will assume that $\Lambda$ can be expressed by $\Lambda^2 = \beta M_{V}^2 + \gamma |t| $, with $\beta = 1$ and $\gamma = 0$. In order to consider the possibility of the production of two different mesons we will generalize this expression for  $\Lambda^2 = \beta_1 M_{V_1}^2 + \beta_2 M_{V_2}^2 + \gamma |t| $. In Fig. \ref{fig2} we analyze the  dependence of the differential cross section for $\rho J/\Psi$ production in our choice for the $\beta_i$ parameters. Although we only present the results for $\gamma \gamma$ collisions at $W = 500$ GeV, the dependence is similar for other center of mass energies. As $\Lambda$ is present in the definition of the variable $z$ [Eq. (\ref{zdef})], which is associated with the energy dependence of the cross section, we have that a different prescription for  $\Lambda$ basically modify the normalization of the differential cross section.  In that follows we will assume $\beta_1 = \beta_2 = 1/2$.

\begin{figure}[t]
\centerline{\psfig{file=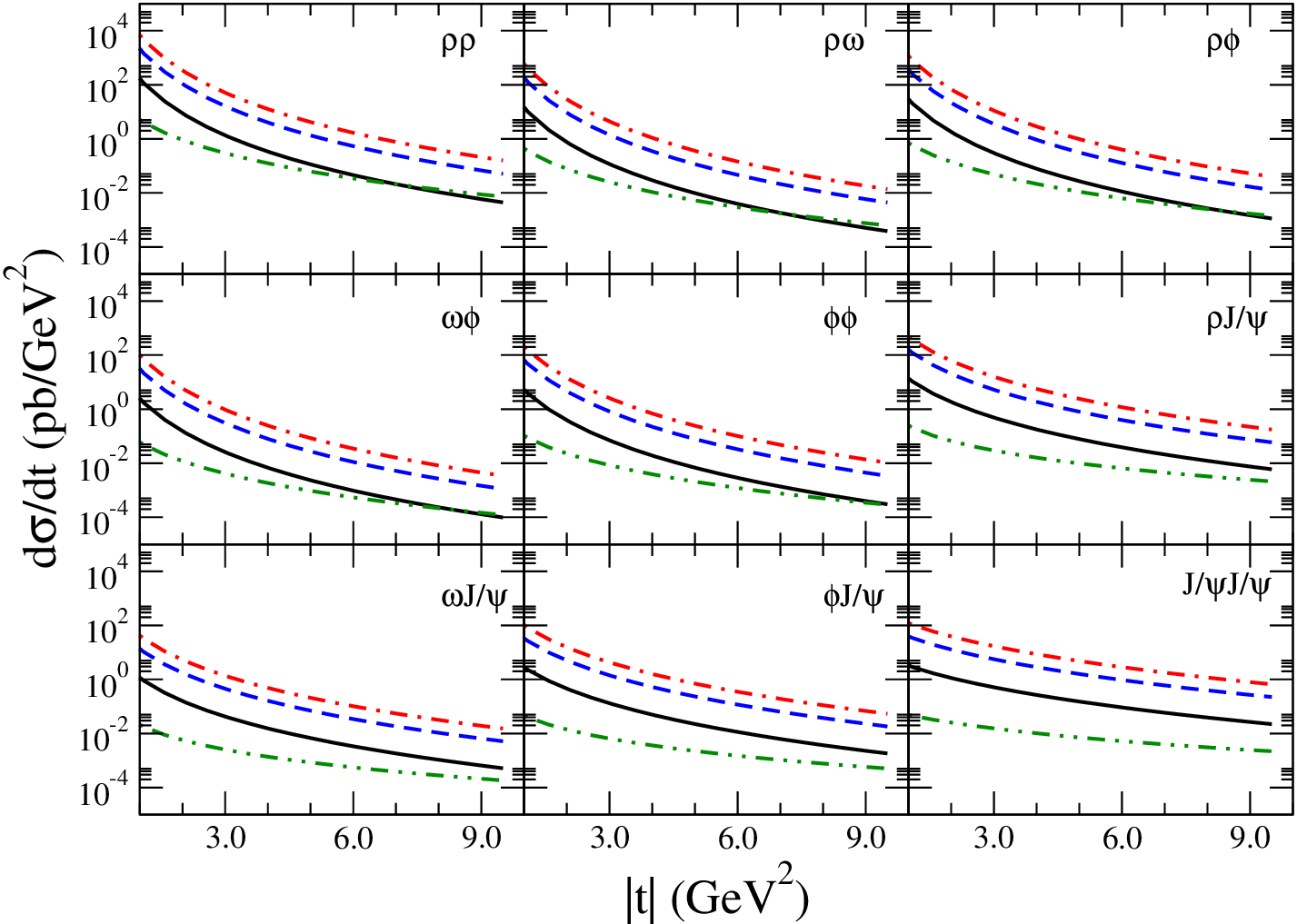,width=125mm}}
 \caption{Momentum transfer dependence of the differential cross sections for the production of  different combinations of pairs of vector mesons and distinct values of center of mass energy. Solid line: $W = 100$ GeV; dashed line: $W = 500$ GeV; dot-dashed line: $W = 1000$ GeV; dot-dot-dashed: Born level.}
\label{fig3}
\end{figure}

In Figs. \ref{fig3} and \ref{fig4} we present our results for the momentum transfer dependence of the differential cross sections for the production of different combinations of pairs of vector mesons and distinct values of energy. For comparison, the Born level prediction which corresponds a two gluon exchange, given by the following formula,
\begin{equation}
\mathcal{F}_{\mathrm{Born}}(s,t) = 2\pi t^2 \int \! \frac{d^2k}{(2\pi)^2}\;\frac{\mathcal{I}_{\gamma V_1}\;\mathcal{I}_{\gamma V_2}}{k_\perp^2 (k_\perp-Q_\perp)^2}
\end{equation}
 is also presented. We have that the differential cross sections decreases  when the masses of the pair of mesons increases as well as it becomes  flatter. Another remarkable feature is the change of the behavior of the cross section when we take into account the BFKL machinery  in comparison with the Born level cross section. The BFKL dynamics implies a growth of the differential cross section and a different slope in comparison to Born level, which is not modified when the energy increases.
As already pointed out in Ref. \cite{motyka}, the inclusion of BFKL dynamics result in a steeper $t$-dependence than the Born calculation.

The total cross section for different combinations of mesons are obtained directly from Eq. (\ref{sigtot}) and the results are displayed in the Figs. \ref{fig5}, \ref{fig6} and \ref{fig7} and in the Tables \ref{tab1} and \ref{tab2}. For light vector meson production, a lower cut-off in the $t$-integration is necessary in order to minimize non-perturbative (Soft Pomeron) contributions. We choose in this case $|t|_{min}=1$ GeV$^2$, which is reasonable considering that the HERA data for the photoproduction of light vector mesons in this kinematic region are  quite well described using a similar approach \cite{FP,jhep}.  
On the other hand, we assume $|t|_{min}=0$ when at least a heavy meson is produced, since in this case we have a  hard scale present which justifies the perturbative calculations in the low-$t$ ($t \approx 0$) region. The predictions obtained using $|t|_{min}=1$ GeV$^2$ are also presented
 in  Figs. \ref{fig5} and \ref{fig6}. We have that  the total cross sections are smaller for heavier vector mesons. Moreover,  the introduction of a cut-off in the $t$-integration implies  a reduction of the normalization of the total cross section, as expected from the  analyses of the differential cross section.
Another important feature is that  difference between the calculations assuming  $|t|_{min}=0$ and $|t|_{min}=1$ GeV$^2$ is  also smaller  for heavier vector mesons. 
In Fig. \ref{fig7} we present our predictions for the production of two light vector mesons assuming $|t|_{min}=1$ GeV$^2$. We have that all cross sections present a strong growth with the energy ($\sigma_{tot} \propto W^{\lambda}$), with a similar slope ($\lambda \approx 1.4$).

In the Tables \ref{tab1} and \ref{tab2} we present a comparison between our predictions for the total cross section for some  values of center of mass energy and the energy independent prediction which comes from the Born amplitude.
We have that the inclusion of the BFKL dynamics implies that the cross sections strongly increases  with the energy, resulting in an  enhancement of the cross section in comparison with the Born level prediction.
This enhancement is larger when a heavy vector meson is produced and $|t|_{min}=0$  GeV$^2$. If we assume $|t|_{min}=1$ GeV$^2$ the BFKL dynamics predicts a similar enhancement of the light and heavy vector meson cross sections.

Lets now compare our results with those obtained in other approaches \cite{ginzburg,motyka,per1,ddr,felix,motykaziaja,double_meson} (For related discussions see Refs. \cite{gregores,pire,rosner}). In Ref. \cite{felix} the  FELIX collaboration has proposed 
the construction of a full acceptance detector for the LHC, with a primary proposal  providing comprehensive observation of a very broad range of strong-interaction processes. In particular, studies of two photon physics in $AA$ collisions has been discussed in its proposal \cite{felix}. There, the differential cross-section $d\sigma (\gamma \gamma \rightarrow V_1 V_2) / dt$, with $V_i = \rho, \, J/\Psi\, ...$,  was parameterized in the form
\begin{eqnarray}
\frac{d\sigma}{dt} (\gamma \gamma \rightarrow V_1 V_2) = A_{V_1 V_2} \, \left(\frac{W}{W_0}\right)^{C_{V_1 V_2}} \, \exp \left(t\, B_{V_1 V_2} + 4 \,t \alpha^{\prime}_{V_1  V_2} \ln \frac{W}{W_0}\right) \,\,. \label{felixeq}
\end{eqnarray}
 This parameterization was obtained assuming the validity of the pomeron-exchange factorization at high energies \cite{gribov} (See also Ref. \cite{block})  which relates the cross sections for different processes. The parameters $A_{V_1 V_2}, \,B_{V_1 V_2} \,, C_{V_1 V_2}$ and   $\alpha^{\prime}_{V_1 V_2} $ were fixed considering the data for vector meson photoproduction at HERA and for $pp$ elastic scattering. For the $\rho J/\Psi$ case, for instance,  $A_{\rho J/\Psi} = 1.1$ nb/GeV$^{-2}$, $B_{\rho J/\Psi} = 2.5$  GeV$^{-2}$, $C_{\rho J/\Psi} = 0.8$ and $\alpha^{\prime}_{\rho J/\Psi} = 0$.
We have assume this parameterization and calculated the total cross section considering $|t|_{min}=0$ 
and $|t|_{min}=1$ GeV$^2$ (See Fig. \ref{fig5}).  In comparison with the BFKL predictions we have that the FELIX parameterization implies a different normalization and energy dependence, with the BFKL one being steeper. Basically, the FELIX parameterization overestimate the cross sections at low energies and underestimate the cross section at very high energies. The only exception occurs when we consider the double $J/\Psi$ production. In this case  the BFKL prediction is larger than the FELIX one for all energies if we assume  $A_{J/\Psi J/\Psi} = 3.1 \times 10^{-4}$ nb/GeV$^{-2}$, $B_{J/\Psi J/\Psi} = 1.5$  GeV$^{-2}$, $C_{J/\Psi J/\Psi} = 1.38$ and $\alpha^{\prime}_{J/\Psi J/\Psi} = 0$. On the other hand, if we assume $A_{J/\Psi J/\Psi} = 0.16/ \ln(W/M_{J/\Psi})$  nb/GeV$^{-2}$ and $C_{J/\Psi J/\Psi} = 2.0$ as also proposed in Ref. \cite{felix} (BFKL-fit) we have that the FELIX parameterization implies a larger cross section for all energies. The main conclusion of the comparison between the BFKL predictions and those obtained assuming the pomeron-exchange factorization is that the BFKL dynamics implies a violation of this factorization, which could be tested in future colliders.

In Ref. \cite{ddr} the double vector meson production in $\gamma \gamma$ collisions has been analyzed using the stochastic vacuum model \cite{vac}. This model is based on the assumption that the infrared behavior of QCD can be approximated by a Gaussian stochastic process in the gluon field strength tensor. A shortcoming of this model is that the energy dependence is not predicted, which implies the introduction of a phenomenological ansatz for this dependence. In  Ref. \cite{ddr} the authors have considered the two-Pomeron model \cite{dl}, resulting that the scattering amplitude is given by  the general expression
\begin{equation}
{\cal A}(W^2, Q^2, t) = \beta_{soft} (Q^2,t) (W^2)^{\alpha_{soft}(t)} +  \beta_{hard} (Q^2,t) (W^2)^{\alpha_{hard}(t)} \,\,,
\end{equation}
where ${\alpha_{soft}(t)} = 1.08 + 0.25 \,t$ and ${\alpha_{hard}(t)} = 1.28$. These values are used in order to describe the $pp$ and $ep$ data. It is important to emphasize that as this model includes soft and hard contributions  the light and heavy vector meson total cross sections can be calculated assuming $|t|_{min}=0$  GeV$^2$. Consequently, its results  for double light vector meson production cannot be directly compared with our predictions. However, analyzing only the energy dependence of both predictions, we have that the BFKL dynamics implies a steeper dependence as expected, since the soft Pomeron contribution dominates the cross sections for these processes in the model proposed in \cite{ddr}.
 On the other hand, when a heavy vector meson is present a comparison can be made. In this case the predictions agree  
for low values of energy ($W \approx 10$ GeV) but are distinct for larger energies  due to the larger intercept resulting from the BFKL equation than the phenomenological one ${\alpha_{hard}(t)}$.

The $\rho J/\Psi$ production in $\gamma \gamma$ processes has also  been estimated in Refs.\cite{motykaziaja,double_meson}. There,
the differential cross section was estimated in  a similar way as the elastic $J/\Psi$ photoproduction off the proton \cite{Ryskin} and reads as \cite{motykaziaja},
\begin{eqnarray}
\frac{d \sigma \,(\gamma \gamma \rightarrow \rho\, J/\Psi)}{dt}\,(W_{\gamma \gamma}^2,t) = {\cal C}\,\,\alpha_{em}\,g_{\rho}^2 \,\frac{16\,\pi^3 \,[\alpha_s(M_{J/\Psi}^2/4)]^2 \,\,\Gamma_{ee}^{J/\Psi}}{3\,\alpha_{em}\,M_{J/\Psi}^5}\, [\,xG^{\rho}(x,M_{J/\Psi}^2/4)\,]^2\,\, \exp\left( B_{\rho\,J/\Psi}\,t\right)\,\,,
\label{sigmat}
\end{eqnarray}
where ${\cal C}$ denotes factors of corrections discussed in detail in Refs. \cite{motykaziaja,double_meson} and  $x=M_{J/\Psi}^2/W_{\gamma \gamma}^2$.  In the small-$t$ approximation, the slope is estimated to be $B_{\rho\,J/\Psi}=5.5 \pm 1.0$ GeV$^{-2}$. The light meson-photon coupling is denoted by $g_{\rho}^2=0.454$. 
The process above was proposed as a probe of the gluon distribution on the meson $xG^{\rho}$ and, by consequence, a constrain for the photon structure. The enhancement in the sensitivity  by taking the square of those distributions in the total cross section  could discriminate them in measurements at the future photon colliders. Considering different models for the gluon distribution of the meson the total cross section was estimated as being  almost 10 pb at $W \approx 10$ GeV and in the range between $100$ and $1000$ pb at $W \approx 100$ GeV. Our results agree with these predictions, with a behavior similar  to those obtained using the GRS(LO) parameterization for the gluon distribution on the  $\rho$ meson (For details see  Refs. \cite{motykaziaja,double_meson}). 
Moreover, in Ref. \cite{double_meson} the double $\rho$ production was  estimated assuming the pomeron-exchange factorization to relate distinct cross sections, similarly to FELIX collaboration.  Our results agree with its results. However, as its predictions are strongly dependent on the assumptions present in the calculations of the double $J/\Psi$ and $\rho J/\Psi$ production (See Table 1 in Ref. \cite{double_meson}), a direct comparison is not very illuminating.     

The first computations of the  double vector meson production at arbitrary momentum transfer were done in Refs. \cite{ginzburg} considering the Born graph for the Pomeron (two-gluon exchange). The total cross section for double light meson production were computed assuming $|t|_{min}=3$ GeV$^2$ and $\alpha_s = 0.32$. Taking into account these differences we have that our calculations of the Born level reasonably agree with those presented in Refs. \cite{ginzburg}. The  double $J/\Psi$ production has been estimated in Refs. \cite{ginzburg,motyka,per1} at Born level and assuming the BFKL dynamics.
Similarly to light vector meson production, our results agree with the predictions from \cite{ginzburg} if the differences are taken into account. In Ref. \cite{per1} the total cross section was estimated assuming a small-$t$ approximation and an ansatz for the $B_{J/\Psi J/\Psi}$ slope parameter. Moreover, it was calculated considering the solution of the forward BFKL solution and some of its next-to-leading corrections. The normalization of the cross section is strongly dependence of the $\alpha_s$ value (See Table 1 in Ref. \cite{per1}) and its energy dependence is steeper when the leading order (LO) solution of the BFKL equation is used. Basically,  although  the normalization of the distinct predictions differ, the energy dependence  is very similar. The difference in the normalization can be explained in terms of the uncertainty introduced in the calculations by the small-$t$ approximation used in Ref. \cite{per1}. On the other hand, if we compare our results with those obtained using the solution of the forward BFKL equation with a different intercept, we have that the normalization for low energies is similar, but the energy dependence of those predictions are softer. This result is expected, since the next-to-leading corrections for the BFKL equation implies a lower intercept than the leading order solution used in our calculations.

Finally, lets estimate the expected number of events of some of the processes calculated in this paper for the future linear colliders.
Currently, the international High Energy Physics (HEP) community is considering the construction of a new $e^+ e^-$ collider \cite{tesla,snow,clic}.  For $\gamma \gamma$ collisions with center of mass energies equal to $W = 500$ GeV, luminosities of order ${\cal{L}} = 1.1$, 0.2 and 4.0 $\times 10^{34} \, cm^{-2} s^{-1}$ are expected at TESLA, CLIC and ILC, respectively. 

It is important to emphasize that the next generation of linear collider can reach values of  center of mass energies up to 3000 GeV. As the BFKL dynamics predicts a strong growth of the cross sections with the energy, the number of events presented in that follows may be considered a lower bound.  In Table \ref{tab3} we present our predictions for the number of events per second for the $\rho \rho$,  $\rho J/\Psi$,  $J/\Psi J/\Psi$ and $\Upsilon \Upsilon$ production. 
For double $\rho$ production our estimate is conservative, since the soft Pomeron contribution is not included in our calculations.
We predict a large number of events related to double meson production in $\gamma \gamma$ collisions, allowing future experimental analyses, even if the acceptance for  vector meson detection were low. Consequently, we believe that this process could be used to constrain the QCD dynamics at high energies.

\section{Summary}
\label{conc}

The photon colliders offer unique possibility to probe QCD in a new and hitherto unexplorated regime. Moreover, good knowledge and understanding of two photon processes will be essential for controlling background contributions to other processes.
In this paper we have studied the double vector meson production using  the non-forward  solution of the BFKL equation  at high energy and large momentum transfer. The total and differential cross section for the process $\gamma \gamma \rightarrow V_1 V_2$, where $V_1$ and $V_2$ can be any two vector mesons  ($V_i = \rho, \omega, \phi, J/\Psi, \Upsilon$), were estimated and compared with those obtained using different approaches. In particular,  we have verified that the pomeron-exchanged factorization leads to different predictions than those obtained in this paper.
 Our main conclusion is that  the forthcoming photon colliders could  experimentally to check our predictions. We believe that this process could be used to constrain the QCD dynamics at high energies. However, several points deserves  more detailed studies: the contribution of all conformal spins, the helicity flip of the quarks, the next-to-leading order corrections to the BFKL dynamics and the corrections associated to the saturation effects.  We plan to estimate  these contributions in future publications.

\section*{Acknowledgments}
The authors are grateful to Magno Machado (UERGS) for illuminating discussions on subjects related to this paper.  W. K. S. thanks the support of the High Energy Physics Phenomenology Group at the Institute of Physics, GFPAE IF-UFRGS, Porto Alegre and the hospitality of IFM-UFPel  where this work was accomplished. This work was partially financed by the Brazilian funding agencies CNPq and FAPERGS.



\newpage

\begin{table}[t]
\begin{center}
\begin{tabular}{||c|r@{.}l|r@{.}l|r@{.}l|r@{.}l||}
\hline
\hline
  & \multicolumn{2}{c|}{Born} & \multicolumn{2}{c|}{$W = 100$ GeV} & \multicolumn{2}{c|}{$W = 500$ GeV} & \multicolumn{2}{c||}{$W = 1000$ GeV} \\
\hline
$\rho \rho$ & 3&18 & 51&0 & 730&0 & 2019&0 \\
\hline
$\rho \omega$ & 0&27 & 4&39 & 59&0 & 173&0  \\
\hline
$\rho \phi$ & 0&48 & 9&52 & 135&0 & 367&0 \\
\hline
$\omega \omega$ & 0&023 & 0&37 & 4&9 & 14&0 \\
\hline
$\omega \phi$ & 0&04 & 0&81 & 11&6 & 31&0 \\
\hline
$\phi \phi$ & 0&076 & 1&87 & 25&2 & 71&0  \\
\hline
\hline
\end{tabular}
\end{center}
\caption{The double vector meson production cross sections in $\gamma \gamma$ processes at different energies and $|t|_{min} = 1$ GeV$^2$. Cross sections are given in pb. } 
\label{tab1}
\end{table}

\begin{table}[t]
\begin{center}
\begin{tabular}{||c| r@{.}l@{~}r@{.}l| r@{.}l@{~}r@{.}l| r@{.}l@{~}r@{.}l| r@{.}l@{~}r@{.}l||}
\hline
\hline
  &  \multicolumn{4}{c|}{Born} & \multicolumn{4}{c|}{$W = 100$ GeV} & \multicolumn{4}{c|}{$W = 500$ GeV}   & \multicolumn{4}{c||}{$W = 1000$ GeV}   \\
\hline
$\rho J/\Psi$ & 1&43 &(0&29)  & 750&0 &(6&8) & 2550&0 &(89&0) & 4833&0 &(234&0)  \\
\hline
$\omega J/\Psi$ & 0&12 &(0&02) & 24&0 &(0&59) & 410&0 &(5&6)  & 1199&0 &(20&0) \\
\hline
$\phi J/\Psi$ & 0&21 &(0&05) & 34&0 &(1&6) & 460&0 &(17&0) & 1647&0 &(55&0) \\
\hline
$J/\Psi J/\Psi$ & 0&19 &(0&10) & 13&0 &(3&6) & 190&0 &(42&0)  & 586&0 &(125&0) \\
\hline
$\rho \Upsilon$ & 0&005 &(0&001) & 0&47 &(0&03) & 6&7 &(0&35) & 24&0 &(1&1) \\
\hline
$\omega \Upsilon$ & 0&0005 &(0&0001) & 0&04 &(0&003) & 0&75 &(0&03) & 2&09 &(0&1) \\
\hline
$\phi \Upsilon$ & 0&001 &(0&0003) & 0&06 &(0&009) & 1&1 &(0&09) & 3&25 &(0&29) \\
\hline
$J/\Psi \Upsilon$ & 0&0025 &(0&0019) & 0&07 &(0&03) & 0&88 &(0&45) & 2&91 &(1&27) \\
\hline
$\Upsilon \Upsilon$ & 0&00013 &(0&00012) & 0&002 &(0&001) & 0&02 &(0&018) & 0&06 &(0&05) \\
\hline
\hline
\end{tabular}
\end{center}
\caption{The double vector meson production cross sections in
$\gamma \gamma$ processes at different energies  assuming $|t|_{min} = 0 \,(1)$ GeV$^2$. Cross sections are given in pb.  } 
\label{tab2}
\end{table}

\begin{table}[t]
\begin{center}
\begin{tabular}{||c|r@{.}l|r@{.}l|r@{.}l||}
\hline
\hline
  & \multicolumn{2}{c|}{TESLA} & \multicolumn{2}{c|}{CLIC} & \multicolumn{2}{c||}{ILC}  \\
\hline
$\rho \rho$  & 8&0 & 2&0 & 28&0  \\
\hline
$\rho J/\Psi$  & 28&0 & 5&0 & 100&0   \\
\hline
$J/\Psi J/\Psi$  & 2&0 & 0&4 & 8&0  \\
\hline
$\Upsilon \Upsilon$  & 0&0002 & 0&00004 & 0&0007 \\
\hline
\hline
\end{tabular}
\end{center}
\caption{Number of events per second for double vector meson production at  TESLA, CLIC and ILC luminosities (W = 500 GeV). For the double $\rho$ production  we assume $|t|_{min}=1$  GeV$^2$. } 
\label{tab3}
\end{table}

\begin{figure}[t]
\centerline{\psfig{file=fig2_dsdt.eps,width=125mm}}
 \caption{Momentum transfer dependence of the differential cross sections for the production of  different combinations of pairs of vector mesons and distinct values of center of mass energy. Solid line: $W = 100$ GeV; dashed line: $W = 500$ GeV; dot-dashed line: $W = 1000$ GeV; dot-dot-dashed: Born level. }
\label{fig4}
\end{figure}

\begin{figure}[t]
\centerline{\psfig{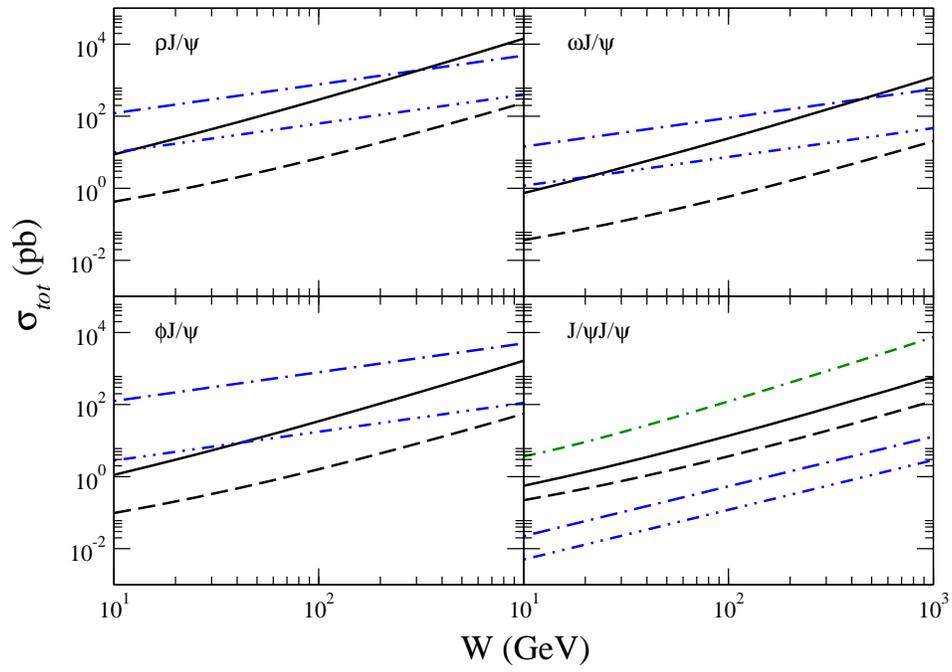}}
 \caption{Total cross section for the production of  different combinations of pairs of vector mesons. Solid line: BFKL ($|t|_{min} = 0$ GeV$^2$); dashed line: BFKL ($|t|_{min} = 1$ GeV$^2$); dot-dashed line: FELIX ($|t|_{min} = 0$ GeV$^2$); dot-dot-dashed line: FELIX ($|t|_{min} = 1$ GeV$^2$). The dot-dashed-dashed line corresponds to the BFKL-fit from FELIX (see text). }
\label{fig5}
\end{figure}

\begin{figure}[t]
\centerline{\psfig{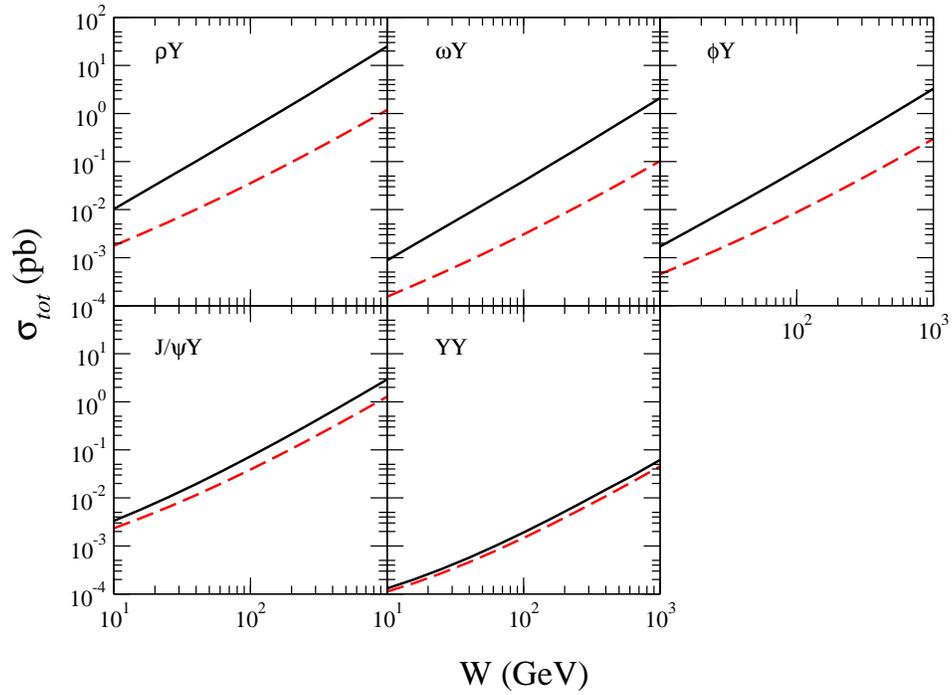}}
 \caption{Total cross section for the production of  different combinations of pairs of vector mesons. Solid line: BFKL ($|t|_{min} = 0$ GeV$^2$); dashed line: BFKL ($|t|_{min} = 1$ GeV$^2$). }
\label{fig6}
\end{figure}

\begin{figure}[t]
\centerline{\psfig{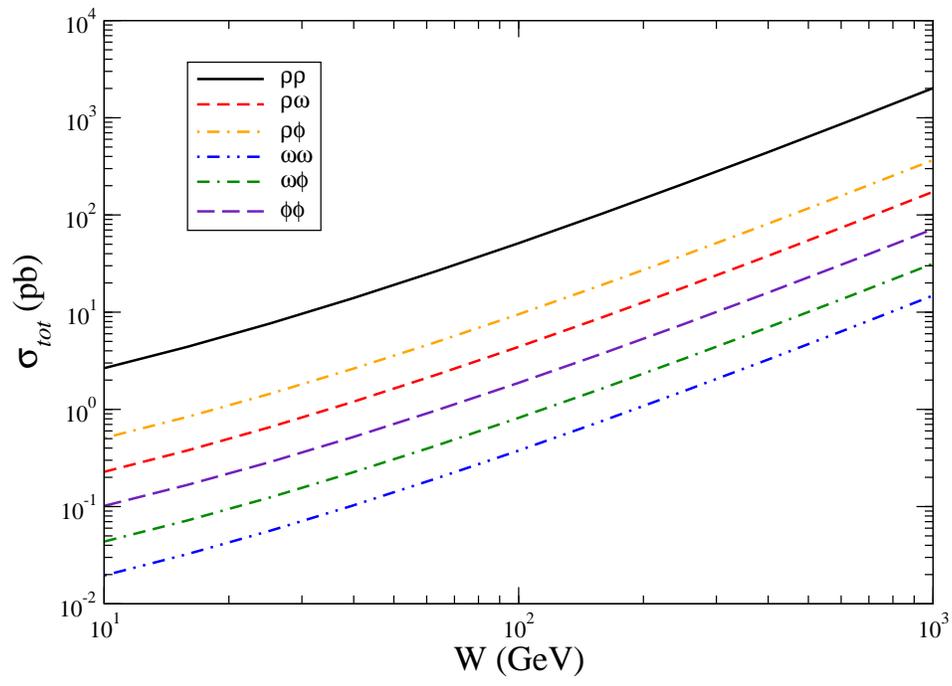}}
 \caption{Energy dependence of the total cross section for the production of  different combinations of pairs of vector mesons ($|t|_{min} = 1$ GeV$^2$). }
\label{fig7}
\end{figure}

\end{document}